\pdfoutput=1

\documentclass[epjST]{svjour}
\usepackage{graphics,amsmath,graphicx}
\begin{document}
\title{Von Neumann's growth model: statistical mechanics and biological applications}
\author{Andrea De Martino\inst{1}
\and Enzo Marinari\inst{2} \and Andrea Romualdi\inst{2}} 
\institute{CNR-IPCF, Dipartimento di Fisica, Sapienza Universit\`a di Roma, Roma, Italy \and Dipartimento di Fisica, Sapienza Universit\`a di Roma, Roma, Italy}
\abstract{
We review recent work on the statistical mechanics of Von Neumann's growth model and discuss its application to cellular metabolic networks. In this context, we present a detailed analysis of the physiological scenario underlying optimality \`a la Von Neumann in the metabolism of the bacterium {\it E. coli}, showing that optimal solutions are characterized by a considerable microscopic flexibility accompanied by a robust emergent picture for the key physiological functions. This suggests that the ideas behind optimal economic growth in Von Neumann's model can be helpful in uncovering functional organization principles of cell energetics.
} 
\maketitle
\section{Introduction}
\label{intro}

In the late 1930's John Von Neumann published (originally in German) a simple linear model to describe optimal economic growth \cite{VN,vn2}. Quite generally, he conceived a system in which $N$ technologies can operate by combining $M$ commodities. Technologies are distinguished by the amounts of goods they use as inputs and return as outputs. Moreover, they are linear (or constant returns to scale in economic jargon), in the sense that doubling the amount of inputs allows to double that of outputs (which implies that, for instance, saturation effects are neglected). Crucially, the system is required to be self-sustained, so that all goods necessary for production at each period must have been produced at the previous period. Given the input/output specifics of production processes, one poses the question of establishing which processes would be operated and which would instead be kept inactive if the system has to maximize the net production rate of commodities. It has been noted that Von Neumann, who had been trained in chemistry and chemical engineering, might have been inspired by chemical reactors for such a construction \cite{brody}. In any case, his growth model has become a cornerstone of the theory of economic growth \cite{gale,dore}. 

Von Neumann's general idea can easily be transported to other contexts outside economics, where input/output processes of various kinds are wired together to form a heterogeneous network \cite{Cohenfw,supply,Fractalriver}. A very notable example is found in cells. Metabolism, namely the complex web of chemical reactions underlying energy transduction in living organisms \cite{Palsson:2006fk,Heinrich:1996uq,Fell:1996kx}, becomes an ideal candidate to test Von Neumann's ideas once the cross-membrane flow of matter is included in the above scheme. Given the fundamental nature of Von Neumann's assumptions, it is natural to ask to which degree the chemical activity of real cells can actually be described by an optimal growth (in the sense discussed above) framework.

On the other hand, the analysis of optimal growth in linear input/output systems is appealing for statistical physicists as well. As the system size increases, the optimal growth properties of `random' input/output networks should hopefully become independent of the particular input-output relationships. Rather, it should be possible to highlight a set of structural characteristics or constraints that are sufficient to frame optimal growth factors and activity profiles. Random input-output systems then constitute a key benchmark against which one can evaluate the performances of single real instances, and the theory of spin glasses and neural networks provides the technical and conceptual tools for this type of analysis \cite{Dotsenko:1995ys}. Indeed, their effectiveness to deal with linear problems close in spirit to Von Neumann's setup  has been demonstrated in various instances in the past (see e.g. the so-called knapsack problem \cite{Korutcheva,Nishimori:2001vn} or even, more recently, the problem of general economic equilibrium \cite{Martino:2004il}).

Our goal here is, on one hand, to look back on the recent work concerned with the statistical mechanics of Von Neumann's growth model for random input-output systems. These studies have revealed a universal behaviour of the relevant quantities and allowed to highlight the key ingredients that can bring the basic framework closer to biologically important questions. Furthermore, we wish to discuss its application to cellular metabolism, which began more recently and is an ongoing and expanding challenge. To complement previous studies, we shall present a broad analysis of the physiological scenario underpinned by Von Neumann's model applied to the cellular metabolism of the bacterium {\it Escherichia coli}. In our view, such studies demonstrate that the basic postulate of optimal growth \`a la Von Neumann might provide a useful key to understand the organization of cell metabolism and energetics.

\section{Von Neumann's model}
\label{sec:1}

The basic setup of Von Neumann's growth model is as follows. One considers a system of $N$ distinct processes (technologies, reactions) that are capable of operating on $M$ compounds (commodities, chemical species). The system is assumed to be closed, i.e. there is no external supply of compounds to the system. In other terms, compounds can only be produced from each other. Each process $i$ is specified by a vector $\mathbf{b}_i=\{b_i^\mu\}$, representing for each $\mu=1,\ldots, M$ the amount of compound $\mu$ process $i$ uses as input (the substrate), and by a vector $\mathbf{a}_i=\{a_i^\mu\}$, representing for each $\mu=1,\ldots, M$ the amount of compound $\mu$ process $i$ outputs (the product). In the simplest case, it is assumed that $a_i^\mu,b_i^\mu\geq 0$ for each $i$ and $\mu$. For sakes of definiteness, each process should have at least one input, whereas each compound should be the output of at least one process. In turn, the input and output vectors form the columns of the $M\times N$ input and output matrices, which we shall denote as $\mathbf{B}$ and $\mathbf{A}$, respectively. Processes are assumed to operate linearly, i.e. when run at scale $s_i\geq 0$ they use amounts $s_i\mathbf{b}_i$ of inputs to produce $s_i\mathbf{a}_i$ outputs. In this setup, the production side (its dual will be discussed later on) of Von Neumann's growth problem (also known as the `technological expansion problem') is formulated in the following terms: 
\begin{equation}\label{max}
\max_{\rho> 0} ~(~\rho~)~~~,~~~~{\rm subject~to~} (\mathbf{A}-\rho\mathbf{B})\mathbf{s}\geq \mathbf{0}
\end{equation}
In other words, one wants to find the largest $\rho>0$ (which we shall denote as $\rho^\star$) such that the system of inequalities
\begin{equation}\label{vn}
c^\mu\equiv \sum_{i=1}^N (a_i^\mu-\rho b_i^\mu)s_i\geq 0~~~\forall\mu
\end{equation}
admits a solution $\mathbf{s}=\{s_i\}$ with $s_i\geq 0$ for all $i$ (with the trivial solution $s_i=0$ being ruled out). The physical meaning of the above conditions becomes clear by noting that the $M$-vector $\mathbf{Bs}$ (resp. $\mathbf{As}$) encodes the total amount of each compound $\mu$ used as input (resp. returned as output) by the various technologies. Hence for fixed $\rho$, a vector $\mathbf{s}$ satisfying (\ref{vn}) guarantees that for each compound the total output is at least $\rho$ times the total input. Maximizing $\rho$ then amounts to finding the largest growth factor sustainable by the system. Note that the growth factor is defined uniformly over compounds, and one does not try to find an optimal vector of growth factors, one for each compound. In such a setting (which would make for an interesting generalization), the $\rho^\star$ defined above would characterize the compounds forming the production bottlenecks, namely those with the smallest feasible growth factor.

In a discrete-time dynamical setup \cite{Martino:2005mi}, one can imagine a scenario in which an operation scale is chosen for every technology at each time step $t$. Then the input vector $\mathbf{I}(t)\equiv \mathbf{Bs}(t)$ generates an output vector $\mathbf{O}(t)\equiv \mathbf{As}(t)$, part of which is recycled into production (i.e. as the input at the next time step). The difference, namely
\begin{equation}
\mathbf{C}(t)=\mathbf{O}(t)-\mathbf{I}(t+1)
\end{equation}
is available e.g. for consumption or for other uses outside the production system strictly defined, including as waste. This condition ensures to be focusing on self-sustaining production patterns. A minimal constraint to be imposed for stability is that $\mathbf{C}(t)\geq 0$, otherwise an external source of compounds would be needed to overcome the input shortages at time step $t+1$. In economics, a central issue at this point is that of maximizing the discounted utility of the overall production over the possible time evolutions. A class of results known as `turnpike theorems' \cite{turnpike} show that optimal paths essentially coincide with the paths of maximal expansion described by Von Neumann's model. Therefore one may restrict to trajectories $\{\mathbf{s}(t)\}$ ensuring that $\mathbf{I}(t+1)=\rho\mathbf{I}(t)$ with $\rho>0$ (i.e. such that $I^\mu(t)=I^\mu(0)\rho^t$). It is easily seen that such paths correspond, for operation scales, to $\mathbf{s}(t+1)=\rho\mathbf{s}(t)$, implying
\begin{equation}
\mathbf{C}(t)=\rho^t(\mathbf{A}-\rho\mathbf{B})\mathbf{s}(0)
\end{equation}
The stability requirement $\mathbf{C}(t)\geq 0$ for all $t$ immediately leads to consider the problem (\ref{vn}) and, in turn, (\ref{max}). It is now clear that if the solution $\rho^\star$ of (\ref{max}) is larger than one the system is expanding (i.e. trajectories are stationary states of constant growth at a growth factor at least equal to $\rho^\star$); if $\rho^\star<1$ the system is instead contracting. 

Note that a linear combination $\lambda\mathbf{s}+\lambda'\mathbf{s'}$ of two solutions $\mathbf{s}$ and $\mathbf{s'}$ with non negative coefficients $\lambda>0$ and $\lambda'>0$ always satisfies the constraints, i.e. the solution space of (\ref{vn}) is convex. As $\rho$ increases (clearly, all vectors $\mathbf{s}$ are solutions for $\rho=0$) one expects its volume to shrink. As we shall see, a key question concerns the size of the solution space that survives at $\rho^\star$.

A classical and easy to prove result concerns the existence of $\rho^\star$ \cite{gale}. Surely, if $\rho$ is sufficiently small the system (\ref{vn}) is bound to admit a solution, i.e. it is possible to find a $\rho>0$ for which $\mathbf{As}\geq\rho\mathbf{Bs}$. On the other hand, it is also possible to choose $\rho>0$ so large that the sum of the row entries of the matrix $\mathbf{A}-\rho\mathbf{B}$ is negative for each row, which would imply the absence of a solution of (\ref{vn}). $\rho^\star$ then coincides with the least upper bound of the set of $\rho$'s for which (\ref{vn}) admits a solution. Establishing the value of $\rho^\star$ is certainly the first task to be carried out. An equally important challenge is however that of profiling the way in which processes are employed in states of optimal growth. Because not all processes need to be operated, one expects that some of them will actually be employed, whereas others will be shut down being unnecessary. Finally we shall focus on the characterization of the set of compounds that will become available for consumption (if any).

\section{Random input-output networks}

The rationale for studying random input-output networks is simply that the macroscopic behaviour of a system of many interacting entities (technologies in Von Neumann's case) can be understood at the statistical level when the number of degrees of freedom diverges. In this limit, in fact, the key macroscopic observables are expected to become substantially independent of the specific realization of the couplings \cite{wigner}. Large systems with random couplings are therefore expected to be good approximations of non-random ones. In this sense, they also constitute a fundamental benchmark against which real systems should be compared. Luckily, a host of analytical and conceptual tools are available to deal with the statistical mechanics of systems with random interactions of the type relevant for Von Neumann's problem \cite{Dotsenko:1995ys,Nishimori:2001vn}.

\subsection{Fully connected networks}

The fully connected Von Neumann model is defined by \cite{Martino:2005mi}
\begin{eqnarray}
a_i^\mu=\overline{a}(1+\alpha_i^\mu)\label{qd1}\\
b_i^\mu=\overline{b}(1+\beta_i^\mu)\label{qd2}
\end{eqnarray}
where $\overline{a}$ and $\overline{b}$ are constants (both $>0$) whereas $\alpha_i^\mu$ and $\beta_i^\mu$ are independent, identically distributed quenched Gaussian random variables of mean zero and variances $\overline{(\alpha_i^\mu)}^2$ and $\overline{(\beta_i^\mu})^2$, respectively. In brief, processes in such a system employ every compound in a finite amount  to produce every compound in a finite amount. The topology underlying this problem is hence that of a fully connected bipartite graph. Inserting the above definitions into (\ref{vn}) one arrives at the conditions
\begin{equation}
(\overline{a}-\rho\overline{b})\sum_{i=1}^N s_i+\sum_{i=1}^N s_i(\overline{a}\alpha_i^\mu-\rho\overline{b}\beta_i^\mu)\geq 0~~~~~\forall\mu
\end{equation}
Because $s_i>0$, for $N\gg 1$ the first term dominates over the second term, which scales like $\sqrt{N}$ (assuming no correlations between $s_i$'s and the Gaussian disorder). Therefore, to leading order in $N$,
\begin{equation}
\rho^\star=\overline{a}/\overline{b}
\end{equation}
This tells us that the leading part of $\rho^\star$ is independent of the details of the input/output vectors. Rather, it is determined only by the average input and output coefficients. For sakes of simplicity, we shall henceforth set $\overline{a}=\overline{b}=1$. To account for subleading corrections, it is convenient to set
\begin{equation}
\rho=1+\frac{g}{\sqrt{N}}
\end{equation}
so that, for $N\gg 1$, $\rho\simeq e^{g/\sqrt{N}}$ and $g$ can be interpreted as a growth rate.  In turn, (\ref{vn}) yields the conditions
\begin{equation}\label{cimu}
c^\mu\equiv \sum_{i=1}^N s_i\left[\alpha_i^\mu-\left(1+\frac{g}{\sqrt{N}}\right)\beta_i^\mu-\frac{g}{\sqrt{N}}\right]\geq 0~~~~~\forall \mu
\end{equation}
One is interested in studying the largest $g$ for which (\ref{cimu}) has solutions. The calculation, detailed in \cite{Martino:2005mi} follows standard steps of spin-glass-like systems (specifically, it is similar to a Gardner calculus \cite{Gardner:1988fk}) and we limit ourselves to a brief sketch here.

The solution space volume at fixed disorder and fixed $g$ can be expressed as
\begin{equation}\label{partfun1}
V(g)={\rm Tr}_{\mathbf{s}}\prod_\mu\theta(c^\mu)\delta\left(\sum_{i=1}^N s_i-N\right)
\end{equation}
where a hard constraint $\sum_{i=1}^N s_i=N$ has been included so as to remove the linear degeneracy inherent in Von Neumann's problem (this is equivalent to setting a scale for the $s_i$'s). The typical solution space volume $V_{typ}(g)\simeq \exp(Nv(g))$ can be obtained from
\begin{equation}
v(g)=\lim_{N\to\infty}\frac{1}{N}\overline{\log V(g)}
\end{equation}
where the over-bar denotes an average over the quenched disorder (\ref{qd1},\ref{qd2}). The evaluation of the disorder average can be carried out resorting to the replica trick. The relevant thermodynamic limit in this case is that where both $N$ (number of processes) and $M$ (number of compounds) diverge with a fixed ratio $n=N/M$. Here, after averages are computed, one can obtain an exact expression for $v(g)$ via a saddle-point integration and a replica-symmetric Ansatz. An interesting outcome of the above calculation is that $\rho^\star$ (or more precisely $g^\star$) depends on the disorder statistics only via the parameter
\begin{equation}
k=\overline{(\alpha_i^\mu-\beta_i^\mu)^2}
\end{equation}
Specifically, one obtains that $g^\star/\sqrt{k}$ is a function of $n$ only. This implies that, generically, larger growth rates require larger spread in the input/output coefficients, suggesting that optimization (and thus process selection) can be operated by choosing maximally diversified technologies. It turns out that for $g\to g^\star$ $v(g)$ can be expressed in terms of the average Euclidean distance between solutions of (\ref{vn}) (denoted as $\chi\equiv\chi(g)$). In particular, one finds
\begin{equation}
v(g)\simeq \chi^{\gamma N}
\end{equation}
where $\gamma$ is a positive exponent. Optimal growth (i.e. $g\to g^\star$) can be retrieved in the limit $\chi\to 0$, in which case $v(g)$ reduces to a single point (i.e. there is a a single configuration of operation scales that corresponds to an optimally growing system). The resulting $(n,g)$ phase diagram is shown in Fig. 1.
\begin{figure}
\begin{center}
\includegraphics[width=8cm]{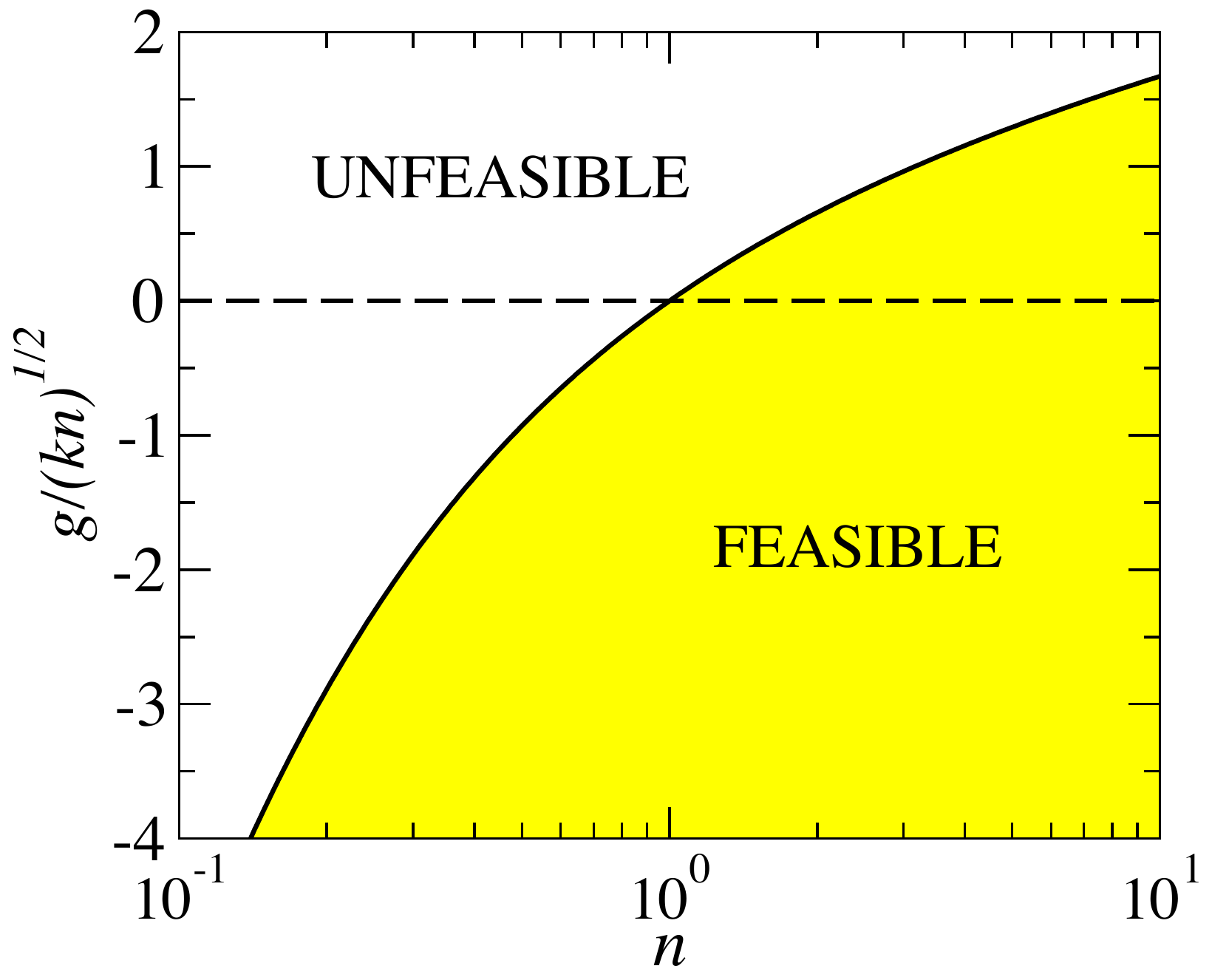}
\caption{Phase structure of the random fully connected Von Neumann model. Values of $n$ and $g$ in the shaded region are viable, i.e. there exist configurations of $s_i$'s satisfying (\ref{vn}). The continuous line is $g^\star/\sqrt{kn}$. For $n<1$ expanding states are unfeasible. Only for $n>1$ can one obtain a feasible expanding state \cite{Martino:2005mi}.}
\end{center}
\end{figure}
In brief, $g^\star<0$ for $n<1$, implying that all viable configurations of operation scales are contracting. When $n>1$, instead, expanding states become viable. The growth rate generically increases with $n$. However the rate of increase changes drastically between the phase where optimal states are contracting to that where expansion is possible. Indeed one gets
\begin{equation}
\frac{\delta\rho}{\delta n}\simeq -\frac{g}{n^{3/2}\sqrt{M}}
\end{equation}
which suggests that an increase in $N$ (i.e. in the number of available technologies) can benefit growth for sufficiently small $n$. By contrast, when $n$ is large technological innovation appears to bear a much smaller effect on the growth properties. The saddle point conditions also allow to derive a simple algebraic equation valid at $g^\star$ that relates the fraction $\psi_0$ of inactive processes (i.e. processes such that $s_i=0$) to the fraction $\phi_0$ of compounds $\mu$ such that $c^\mu=0$ (i.e. for which the overall output exactly matches the overall input):
\begin{equation}\label{alg}
\psi_0=1-\phi_0/n
\end{equation}
In different terms, (\ref{alg}) simply states that for $g=g^\star$ (when a single vector of operation scales survives satisfying (\ref{vn})) the number of variables $s_i\neq 0$ (namely $N(1-\psi_0)$) should equal the number of equalities (namely $M\phi_0$). Both $\psi_0$ and $\phi_0$ increase with $n$ in a roughly sigmoidal way, with $\psi_0(n=1)=\phi_0(n=1)=1/2$. In words, optimal states in the contracting regime ($n\ll 1$) are characterized by having  almost all processes operated and almost all compounds available for consumption. In optimal expanding states, instead ($n\gg 1$), a small fraction of processes is profitable while almost all compounds are perfectly recycled with limited production of consumable goods or waste.

\subsection{Finitely connected networks}

A mathematically more compelling scenario is obtained when the topology underlying the production networks is characterized by finite connectivity, i.e. when each process uses a finite number of compounds as substrates to produce a finite number of other (different) compounds \cite{Martino:2007zt}. This more realistic case can be conveniently dealt with by the cavity method under the assumption that the bipartite graph that underlies the production network has a locally tree-like structure. In essence, cavity theories exploit the fact that on such graphs the removal of a node of one type (either a compound or a reaction) makes the nodes of the other types (reactions or compounds) that were linked to the node we removed roughly statistically independent, since short loops causing feedbacks and correlations are assumed to be absent (or anyway negligible). So if we denote by $\mathbf{s}_{i\in \mu}$ the scale variables of processes connected to a compound $\mu$ and by $\mathbf{c}^{\mu\in i}$ the variables defined in (\ref{vn}) for compounds connected to a process $i$, then their probability distributions are such that
\begin{equation}
p(\mathbf{s}_{i\in \mu})\neq \prod_{i\in\mu}p_i(s_i)~~~{\rm and}~~~
q(\mathbf{c}^{\mu\in i})\neq \prod_{\mu\in i} q^\mu(c^\mu)
\end{equation}
unless compound $\mu$ or, respectively, process $i$ are removed, in which case one has
\begin{equation}
p^{(\mu)}(\mathbf{s}_{i\in \mu})= \prod_{i\in\mu}p_i^{(\mu)}(s_i)~~~{\rm and}~~~
q_{(i)}(\mathbf{c}^{\mu\in i})=\prod_{\mu\in i} q_{(i)}^\mu(c^\mu)
\end{equation}
where the subscripts $(i)$ and $(\mu)$ indicate the removal of process node $i$ and of compound $\mu$, respectively. Note however that $q_{(i)}^\mu(c^\mu)$ can be written explicitly as
\begin{equation}\label{cav2}
q_{(i)}^\mu(c^\mu)={\rm Tr}_{\mathbf{s}_{j\in\mu\setminus i}}
\delta\left(c^\mu-\sum_{j\in\mu\setminus i}s_j(a_j^\mu-\rho b_j^\mu)\right)
\prod_{j\in\mu\setminus i}p_j^{(\mu)}(s_j)
\end{equation}
where $j\in\mu\setminus i$ indicates that the operations extend over all processes connected to $\mu$ except $i$. Similarly, introducing the fictitious energy function
\begin{equation}
E=-\sum_{\mu=1}^M\theta(c^\mu)
\end{equation}
one has
\begin{equation}\label{cav1}
p_i^{(\mu)}(s_i)\propto {\rm Tr}_{\mathbf{c}^{\nu\in i\setminus\mu}}
\exp\left[-\beta\sum_{\nu\in i\setminus\mu}\theta[-c^\nu_{(i)}-s_i(a_i^\nu-\rho b_i^\nu)]\right]
\prod_{\nu\in i\setminus\mu} q_{(i)}^\nu(c^\nu)
\end{equation}
(ultimately, the limit of large $\beta$ must be considered). In turn, knowledge of the so-called cavity distributions (\ref{cav2}) and (\ref{cav1}) allows to reconstruct the entire probability distribution since
\begin{gather}
p_i(s_i)\propto {\rm Tr}_{\mathbf{c}^{\nu\in i}}q_{(i)}(\mathbf{c}^{\nu\in i})\exp
\left[-\beta\sum_{\nu\in i}\theta[-c^\nu_{(i)}-s_i(a_i^\nu-\rho b_i^\nu)]\right]\\
q^\mu(c^\mu)={\rm Tr}_{\mathbf{s}_{j\in\mu}} p^{(\mu)}(\mathbf{s}_{j\in\mu})
\delta\left(c^\mu-\sum_{j\in\mu}s_j(a_j^\mu-\rho b_j^\mu)\right)
\end{gather}
Equations (\ref{cav2}) and (\ref{cav1}) can be solved self-consistently by population dynamics. The numerical procedure to extract $\rho^\star$ from this setup is non-trivial and we refer the reader to \cite{Martino:2007zt} for details. The lesson one learns from this analysis is that the picture obtained in the fully connected case is rather robust to dilution. Finite connectivity does however bear a strong quantitative impact. In the contracting regime ($\rho^\star<1$) optimal growth rates generically decrease with increasing dilution while the opposite happens in expanding regimes with $\rho^\star>1$. Moreover, optimal growth rates turn out to be consistently larger when processes have heterogeneous (as opposed to regular) connectivities, independently of the degree distribution of compounds. Hence, generically dilution (in expanding regimes) and stochastic connectivities allow to achieve larger growth rates with respect to the case of fully connected input/output systems. The phase structure shown in Fig. 1 is however qualitatively preserved.

\subsection{Reversible networks}

An important generalization considers processes as reversible, i.e. such that the roles of input and output coefficients can be interchanged \cite{Martino:2010tw}. This simple extra ingredient generates a surprisingly rich and considerably more complex scenario. Let us first notice that by allowing processes to revert Von Neumann's problem becomes intrinsically non-linear. Indeed it can be written compactly upon defining a matrix $\boldsymbol{R}$ with entries
\begin{equation}
R_i^\mu=
\begin{cases}
a_i^\mu-\rho b_i^\mu&\text{for $s_i>0$}\\
b_i^\mu-\rho a_i^\mu&\text{for $s_i<0$}
\end{cases}
\end{equation}
whereby one sees that $\mathbf{R}\equiv\mathbf{R}(\rho,\mathbf{s})$. (The fact that process can be reverted is reflected also in the possibility that $s_i$ becomes negative.) Under this condition, (\ref{max}) now reads
\begin{equation}
\max_{\rho> 0} ~\rho~~~{\rm subject~to~} \mathbf{R(\rho,\mathbf{s})s}\geq \mathbf{0}
\end{equation}
There are two basic properties of the reversible Von Neumann problem that are both easy to prove and rich of consequences. The first is that in general reversing all processes is not equivalent to time reversal, i.e. the naive expectation that if a scale vector $\mathbf{s}$ guarantees a growth factor $\rho$ then the reversed vector $\mathbf{s'}$ should guarantee a growth factor $\rho'=1/\rho$ turns out to be false. To see this, define
\begin{eqnarray}
c_d^\mu(\mathbf{s})=\sum_{i=1}^N s_i(a_i^\mu-\rho b_i^\mu)\\
c_r^\mu(\mathbf{s})=\sum_{i=1}^N s_i(b_i^\mu-\rho' a_i^\mu)
\end{eqnarray}
and note that
\begin{equation}
c_d^\mu(\mathbf{s})+\rho c_r^\mu(\mathbf{s})=(1-\rho\rho')\sum_{i=1}^N s_i a_i^\mu
\end{equation}
Clearly, $c_d^\mu(\mathbf{s})=c_r^\mu(\mathbf{s})=0$ implies $\rho'=1/\rho$ but if $c_r^\mu(\mathbf{s})\geq 0$ then
\begin{equation}
\rho\rho'\leq 1-\frac{c_d^\mu(\mathbf{s})}{\sum_{i=1}^N s_i a_i^\mu}
\end{equation}
which tells us that $c_d^\mu(\mathbf{s})=0$ for all $\mu$ is a necessary but not sufficient condition for $\rho'$ to be equal to $1/\rho$. In general,
\begin{equation}
\rho\rho'\leq \min_\mu \left(1-\frac{c_d^\mu(\mathbf{s})}{\sum_{i=1}^N s_i a_i^\mu}\right)
\end{equation}
The second notable fact is that the solution space ceases to be convex, in particular for $\rho<1$. Re-writing (\ref{vn}) in this case as
\begin{equation}
c^\mu(\mathbf{s})=\sum_{i=1}^N s_i \left[a_i^\mu(\theta(s_i)+\rho\theta(-s_i))-b_i^\mu(\theta(-s_i)+\rho\theta(s_i))\right]\geq 0~~~~~\forall\mu
\end{equation}
one easily sees that
\begin{equation}
c^\mu(\lambda\mathbf{s}+\lambda'\mathbf{s'})=\lambda c^\mu(\mathbf{s})+\lambda' c^\mu(\mathbf{s'})+A^\mu(\mathbf{s},\mathbf{s'})
\end{equation}
with
\begin{equation}
A^\mu(\mathbf{s},\mathbf{s'})=(1-\rho)\sum_{i=1}^N(a_i^\mu+b_i^\mu)[\lambda s_i\theta(-s_i)+\lambda' s_i'\theta(-s_i')-(\lambda s_i+\lambda' s_i')\theta(-\lambda s_i-\lambda' s_i')]
\end{equation}
If $\mathbf{s}$ and $\mathbf{s'}$ are both solutions, then a linear combination will again be a solution if $A^\mu\geq 0$. $s_i s_i'>0$ indeed implies $A^\mu=0$ for each $\rho$. However for $s_i s_i'<0$ one finds that $A^\mu\geq 0$ only for $\rho\geq 1$. In other words convexity is guaranteed only for $\rho\geq 1$ and may not occur even at $\rho^\star$ when $\rho^\star<1$. Note however that the solution space is obviously convex for $\rho=0$, showing that $\rho<1$ is a necessary but not sufficient condition for the solution space to be non-convex (e.g. fragmented in disjoint parts). It is very important to remark that non-convexity poses a serious  problem for the design of efficient algorithms to sample the solution space in the contracting regime \cite{Martino:2010tw}.

The statistical mechanics analysis of this case in a fully connected topology can proceed along steps similar to those performed for the random fully connected network, with (\ref{partfun1}) replaced by
\begin{equation}\label{partfun2}
V(g)={\rm Tr}_{\mathbf{s}}\prod_\mu\theta(c^\mu)\delta\left(\sum_{i=1}^N |s_i|-N\right)
\end{equation}
If $\phi$ denotes the fraction of reversible processes (i.e. the effective number of processes that can be operated in the system is $N+N\phi$), one finds an expanding phase with $g^\star > 0$ for $n > n_c\equiv 1/(1 + \phi)$ (see Figure 2). 
\begin{figure}
\begin{center}
\includegraphics[width=8cm]{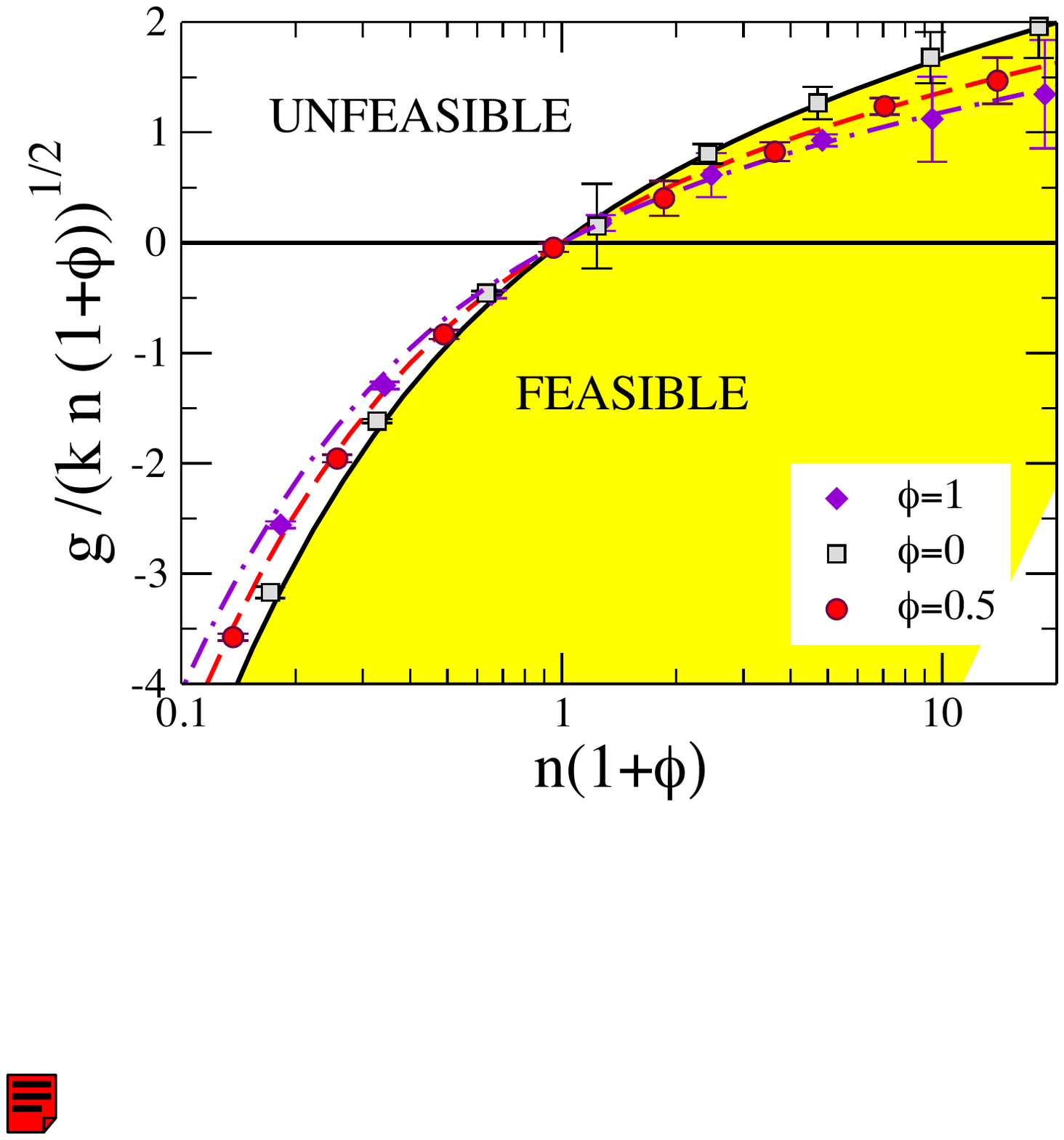}
\caption{Phase structure of the random fully connected reversible Von Neumann model. The continuous (resp. dashed and dot-dashed) line is $g^\star/\sqrt{kn(1+\phi)}$ for $\phi=0$ (resp. $\phi=1/2$ and $\phi=1$). Markers correspond to optimal growth rates estimated numerically as explained in \cite{Martino:2010tw}. (For clarity, the shaded region is only shown for $\phi=0$.) For $n(1+\phi)<1$ expanding states are unfeasible. However one sees slight disagreements between the predicted (replica-symmetric) value of $g^\star$ and computed one.}
\end{center}
\end{figure}
This suggests, as expected, that being able to run processes in both directions extends the range of the regime where expanding solutions exist. The price to be paid is that optimal growth rates in the expanding regime are comparatively smaller when processes are reversible. For $n(1+\phi)<1$, instead, the system is confined to a contracting regime. Here, larger $\phi$'s imply larger growth rates, meaning that increasing the availability of reversible processes benefits the system's growth properties. As seen above, however, the contracting phase is characterized by lack of convexity. This is hinted by the fact that the computed and predicted (replica-symmetric) values of $g^\star$ disagree (albeit just slightly) for $n<n_c$ when $\phi\neq 0$. To see this more explicitly one can measure directly the probability that the uniform linear combination of two solutions solves Von Neumann's problem (Figure 3). 
\begin{figure}
\begin{center}
\includegraphics[width=8cm]{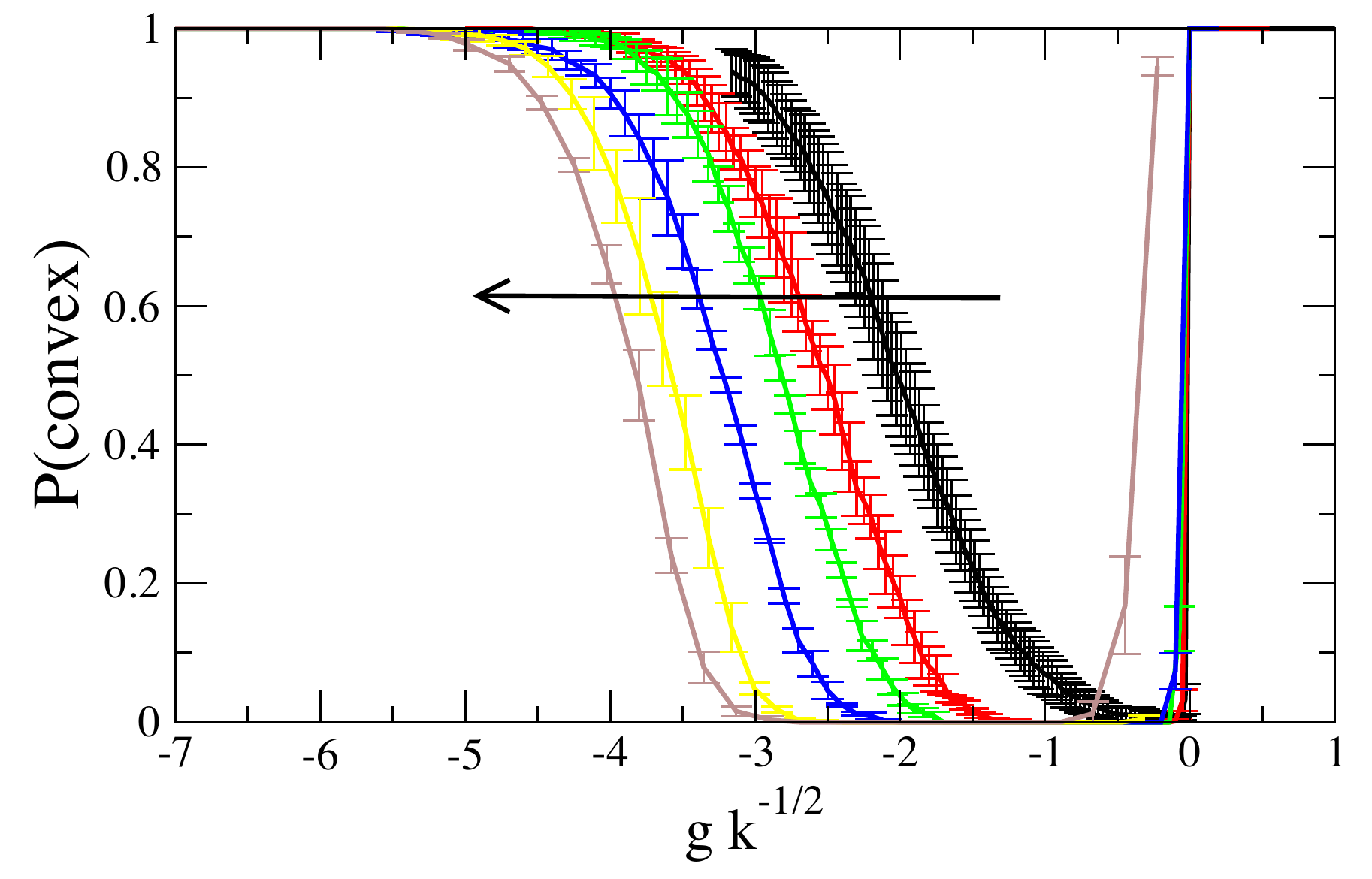}
\caption{Probability that the linear combination of two solutions with coefficients $1/2$  solves the fully reversible ($\phi=1$) random fully connected von Neumann problem for $n=1/2$. The solution space appears to lose convex values of $g<g^\star=0$. Different curves correspond to different values of $N$ increasing in the direction of the arrow from 50 to 500.}
\end{center}
\end{figure}
Inspecting results, one sees immediately that the solution space displays marks of convexity for sufficiently small $g<g^\star$ (recall that the solution space is certainly convex for $\rho=0$). In the intermediate range of values of $g$ up to $g^\star$, instead, convexity appears to break down. The replica-symmetric theory one develops for this case is thus expected to fail in a range of values of $g<0$ (as seen in Fig 2). 

As before, an increase in the optimal growth rate causes a decrease in the number of compounds with $c^\mu>0$. This corresponds to a `waste' reduction and, on the other hand, to an impoverishment in the global output profile. Quite strikingly, the reversible part of the process set displays a neat second-order transition as $n$ exceeds $n_c$: all reversible processes are employed in the contracting phase, where a finite fraction of them is inactive when $g^\star>0$ \cite{Martino:2010tw}.

\section{Towards metabolic networks}

Von Neumann's problem finds a natural ground of application in chemical reaction networks, in which the input-output relations linking technologies to goods in production systems are replaced by the wiring between reactions and chemical species governed by stoichiometry. Perhaps most importantly, in this context living cells present us the opportunity to test Von Neumann's ideas directly in a real system, namely their energy metabolism.

In cells, an enormous number of genetic, regulatory and biochemical mechanisms interact on different spatial and temporal scales to accomplish a wide range of tasks, from growth and reproduction to motility, homeostasis and exterior sensing. The rapid development of high-throughput genome sequencing techniques together with the availability of refined gene annotations has allowed, over the past few years, to set up consistent, controlled protocols for reconstructing the cellular metabolic networks of different organisms to an unprecedented degree of detail \cite{Palsson:2006fk,Thiele:2010bs}. Such networks underlie the basic energy harvesting processes in cells,  by which the energy derived from nutrients is transduced into usable forms of mechanical or chemical energy. The output of metabolism is constituted essentially by the so-called building blocks  for functionally relevant biological macromolecules  (amino-acids, fatty acids, nucleotides) as well as by waste products like e.g. carbon dioxide and acetate (mostly for bacteria). Building blocks are then used in a series of processes downstream of metabolism to form functional proteins (including reaction-catalying enzymes), membrane structures and nucleic acids. To a large degree, the physiological outcome of this can be read off from the cell's proteome, i.e. from the composition of the repertoire of proteins that the organism was able to produce, which in turn effects the regulation of metabolism. 

It is immediately clear that, once the structure of the network (encoded in the reactions' stoichiometry) is known, Von Neumann's stability criterion can serve as a minimal model for the organization of metabolism. However, before discussing this application more in detail (Sec. 5), it is worth trying to bring the theory a few steps forward (in a direction relevant to biology) by addressing a couple of peculiar aspects of biochemical reaction networks that, on one hand, have a direct influence on the solutions of Von Neumann's problem and, on the other, generalize it. These are, respectively, (a) the emergence of the so-called `conserved pools of reagents', which translate into the need to consider constraints on the quenched disorder (the input-output coefficients), and (b) the highly important issue of thermodynamic stability.

\subsection{Stoichiometric quenched disorder}

In first place, addressing optimality in reaction networks requires a more careful evaluation of the role of stoichiometry, since real networks are of course not random: stoichiometric coefficients enforce chemical balance at each reaction node in the network. This set of local relations has been shown to give rise, at a larger scale, to conservation laws for the aggregate concentration of pools of reagents \cite{Famili:2003fk}. Such laws are expressed by the relation 
\begin{equation}
\sum_{\mu\in P}(a_i^\mu-b_i^\mu)=0 ~~~~~\forall i
\end{equation}
where $P\subseteq\{1,\ldots M\}$ is the set containing the chemical species that form the conserved pool. In concrete, the total number of molecules in a pool is conserved over time, so that the aggregate concentration does not change in time (while the concentrations of individual metabolites can change relative to each other). These dynamical invariants have a purely topological origin in the structure of stoichiometric coefficients. As a byproduct, one easily understands that metabolites belonging to a conserved pool cannot be global network outputs (a detailed theory of this fact with applications is developed in \cite{Imielinski:2005dz,ib00}). In other terms, in presence of a conserved metabolite pool one expects that $\rho^\star\leq 1$.

A first step in driving Von Neumann's model towards real cellular networks consists therefore in studying how the existence of conserved metabolite pools affects the picture derived in the case of random networks. The simplest way to implement this ingredient is to impose a constraint on the quenched disorder. Specifically, one can introduce a `random' conserved pool formed by a finite fraction $\phi$ of reagents by explicitly adding to the volume defined in (\ref{partfun1}) a hard constraint of the form
\begin{equation}\label{pool}
\sum_{\mu=1}^M z^\mu(a_i^\mu-b_i^\mu)=0
\end{equation}
(where $z^\mu$ is a quenched random variable equal to one with probability $\epsilon$ and zero otherwise). In order to check explicitly that the constraint (\ref{pool}) implies $\rho^\star\leq 1$, i.e. that at best optimal growth states are characterized by constant operation scales for reactions (fluxes), it suffices to multiply each term in (\ref{vn}) by $z^\mu$ and sum over $\mu$ (we ignore reversibility). One obtains:
\begin{equation}
(1-\rho)\sum_{i=1}^N s_i\sum_{\mu=1}^M z^\mu b_i^\mu+\sum_{i=1}^N s_i\sum_{\mu=1}^M z^\mu(a_i^\mu-b_i^\mu)\geq 0
\end{equation}
If the vector $\mathbf{z}$ identifies a conserved pool then the second term vanishes. The first term allows for $\rho>1$ only if all reactions connected to metabolites in the conserved pool are inactive (i.e. if the pool is effectively removed from the network). In particular this leads to the trivial solution $s_i=0$ for each $i$ in a fully connected network. Hence necessarily $\rho^\star\leq 1$. So the states of expansion recovered in the previous sections are a `pathology' of random systems. When stoichiometric quenched disorder is used they cannot be present. Notice that this does not mean that the network cannot have a positive net production of compounds. The introduction of (\ref{pool}) slightly modifies the disorder averaging but the replica theory developed for random input/output systems can be conveniently extended to this constrained case \cite{Martino:2009lh}. The $(\epsilon, n)$ phase structure one obtains, in agreement with the previous observations, is shown in Figure 4.
\begin{figure}
\begin{center}
\includegraphics[width=8cm]{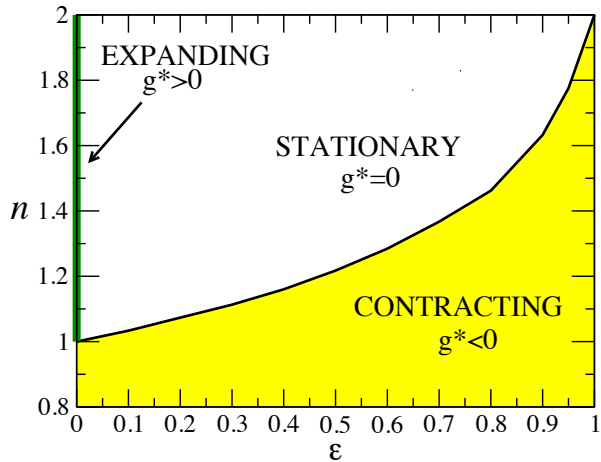}
\caption{Phase structure of the random fully connected constrained Von Neumann model. The continuous line marks the separation between the regions with $g^\star=0$ (stationary regime) and $g^\star<0$ (contracting regime). The expanding regime lies on the $\epsilon=0$ axis ($n>1$) and is represented by a green strip \cite{Martino:2009lh}.}
\end{center}
\end{figure}

A noteworthy consequence of the fact that $\rho^\star$ cannot exceed 1 is that for large enough $n$, when an expanding phase would be expected in absence of the conserved pool, multiple optima survive, i.e. the average distance between solutions does not vanish as $g\to g^\star$. So stoichiometric disorder appears to increase robustness in reaction networks, since many (microscopic) arrangements of reaction operation scales turn out to be compatible with optimal growth. It is interesting to note that real cellular metabolic networks typically have values of $n$ larger than one. This suggest that the ability to respond to perturbations (e.g. by re-arranging scales) without losing the desirable property of maximum productive capacity may be a key ingredient for biological stability and resilience.

\subsection{Von Neumann's dual problem and thermodynamics}

Thus far, we have limited ourselves to considering Von Neumann's problem in a production perspective. Our focus has been set on finding ways to operate the available processes that ensure self-sustainability and optimality. In his original work however J. Von Neumann has given equal attention to the dual problem (in the sense of linear programming duality) \cite{vn2}. In elementary economic terms, it can be introduced upon defining an $M$-vector $\mathbf{p}$ of prices ($p^\mu\geq 0$ for each $\mu$). The cost associated to running a certain process $i$ at unit scale is simply $\sum_{\mu=1}^M p^\mu b_i^\mu$. This generates a corresponding revenue given by $\sum_{\mu=1}^M p^\mu a_i^\mu$. The problem dual to (\ref{max}) consists in solving
\begin{equation}\label{min}
\min_{\sigma> 0} ~\sigma~~~{\rm subject~to~} \mathbf{p}(\mathbf{A}-\sigma\mathbf{B})\leq  \mathbf{0}
\end{equation}
(compare this with (\ref{max})). The interpretation of the parameter $\sigma$ is not straightforward. The simplest is in terms of an interest factor. Imagining that the operation of processes is financed by borrowing cash at a rate $r$, at the end of each period $\sigma=1+r$ units of cash must be returned to the financiers. In this setting, the conditions
\begin{equation}\label{conda}
\sum_{\mu=1}^M p^\mu(a_i^\mu-\sigma b_i^\mu)\leq 0~~~~~\forall i
\end{equation}
simply mean that the net extra profit generated by each process can at most be zero. (Zero profit is a standard scenario in models of economic equilibrium, see e.g. 
\cite{Debreu:1972fk} or, in a statistical physics perspective, \cite{macroec}.)

It is simple to show that, in the above settings of (\ref{max}) and (\ref{min}) and under broad assumptions, $\rho^\star\geq \sigma^\star$ (with $\sigma^\star$ the solution of (\ref{min})) \cite{gale}. Furthermore, restricting slightly the definition of the model it can be proven that indeed $\rho^\star=\sigma^\star$ (i.e. that the growth factor equals the interest factor, as would perhaps be intuitively expectable).

It is however tempting to search for a physical interpretation of Von Neumann's dual problem. The simplest connection resides in the thermodynamic scenario underlying biochemical activity. It is well known that in non equilibrium steady states with non-zero fluxes, reactions must proceed in directions of decreasing free energies (which in turn implies that thermodynamically feasible configurations of reaction fluxes may not contain directed cycles) \cite{Beard:2008fk}. The condition for thermodynamic feasibility reads
\begin{equation}
v_i\Delta G_i\leq 0~~~~~\forall i
\end{equation}
where $\Delta G_i$ is the (Gibbs) free energy change associated to reaction $i$ and $v_i$ is the net flux (the difference between the forward and reverse rates) of reaction $i$. $\Delta G_i$ can be written as the sum of the chemical potentials of the compounds weighted by their respective stoichiometric coefficients. It then follows that thermodynamic feasibility is related to the existence of a chemical potential vector $\mathbf{g}$ such that
\begin{equation}
v_i\sum_{\mu=1}^M g^\mu(a_i^\mu-b_i^\mu)\leq 0~~~~~\forall i
\end{equation}
Connection to (\ref{conda}) is now straightforward. In this respect, prices play the role of chemical potentials and (\ref{conda}) have the form of a thermodynamic feasibility constraint. 

It would be interesting to construct a more stringent physical analogy for (\ref{min}). Quite recently an attempt in this direction has been performed in conjunction with Flux-Balance-Analysis (FBA) \cite{Palsson:2006fk,Orth:2010if}. It was shown in specific that the dual to the linear programming problem that arises in FBA can be re-formulated as a free energy consumption minimization problem conditioned on a given free energy drain (representing, in the case of cellular systems, growth) \cite{Warren:2007fk}.

\section{Application to cellular metabolism}

Modeling a cell's metabolic activity is a central problem in systems biology. In principle, knowledge of the network structure and of basic thermodynamic parameters like reaction constants allows to formulate non-linear differential equations for the enzyme-mediated joint evolution of reactant concentrations and reaction rates (or fluxes) that fully account for the kinetics of a biochemical reaction network \cite{Beard:2008fk}. Unfortunately at genome scale this route is most often barred by uncertainties about kinetic parameters and reaction or transport mechanisms. Constraint-based models of metabolism (like Flux-Balance Analysis \cite{Palsson:2006fk,Beard:2008fk,Orth:2010if,Kauffman:2003ff}) provide a feasible alternative that has marked a considerable breakthrough in the quest to reconstruct and simulate global cellular functions. In a nutshell, they rely on steady state assumptions to impose sets of physically motivated linear constraints (enforcing mass balance) on the network nodes corresponding to reagents, and retrieve the viable flux states as the configurations of reaction operation scales that ensure that  all constraints are satisfied. Regulatory and thermodynamic restrictions are included either in the allowed ranges of variability of fluxes or in the a priori reversibility assignments of reactions. Such a constrained linear system  suffices to define a space of feasible flux configurations (whose characterization is in itself a challenging problem \cite{Braunstein:2008bf,Bianconi:2008km}). Quite crucially, physiologically relevant states are usually assumed to maximize {\it ad hoc} objective functions that provide a further functional constraint on the cell's metabolic activity (see e.g. \cite{Segre:2002oq,Holzhutter:2004pr,Shlomi:2005fk,Schuetz:2007le} for some examples). 

Applying Von Neumann's idea to cellular metabolism amounts to taking a rather different (though not unrelated) viewpoint. Recalling that in real stoichiometric systems $\rho^\star=1$, finding optimal solutions in the Von Neumann sense means finding flux vectors $\mathbf{s}$ such that
\begin{equation}\label{vn2}
c^\mu\equiv \sum_{i=1}^N (a_i^\mu-b_i^\mu)s_i\geq 0~~~\forall\mu
\end{equation}
where the index $i$ runs over reactions, $\mu$ over chemical species and $a_i^\mu$ and $b_i^\mu$ denote the output and input (respectively) stoichiometric coefficients of metabolite $\mu$ in reaction $i$. As explained above, (\ref{vn2}) encodes in essence a stability constraint. If fluxes that transport matter (e.g. nutrients) into the cell are included in the stoichiometric matrices to make the system self-sustainable (in such a way that the nutrient consumption equals its influx), then solving (\ref{vn2}) allows to retrieve information about what the cell is in principle capable of net-producing {\it in a given environment}, as well as on the corresponding flux configurations. Metabolites for which $c^\mu=0$ indeed turn out to be mass-balanced, so that their production and consumption fluxes exactly match. On the contrary, if $c^\mu>0$ then metabolite $\mu$ is globally producible, meaning that it can become available either for macromolecular processes or as waste. The availability of a neural network learning-based algorithm for the efficient generation of solutions of (\ref{vn2}) (see \cite{Martino:2007zt}, based on \cite{minover0}) then permits a detailed statistical analysis of the cell's metabolic capabilities. 

Work performed along these lines for the bacterium {\it Escherichia coli} \cite{Martelli:2009jl,Kyoto10} has shown that environment selection restricts the feasible states according to (\ref{vn2}) to flux configurations that both reproduce well the available empirical evidence on E.coli's metabolic fluxes and guarantee that the correct physiological task (in that case, growth or biomass production) is carried out by the cell without the need of additional {\it ad hoc} functional constraints. We want to add here further evidence supporting the relevance of Von Neumann's observations in the context of cell metabolism.

We have computed and analyzed optimal flux configurations (in Von Neumann's sense) for the {\it Escherichia coli} metabolic network reconstructed in \cite{Reed:2003ij}, preparing it in a minimal growth medium with gluxose as its main carbon source. At odds with the work presented in \cite{Martelli:2009jl,Kyoto10}, we have modified the model to be able to measure the levels of some of the physiological objective functions used in the computational biology literature (see e.g. \cite{Schuetz:2007le}). Our goal is to characterize the solution space defined by (\ref{vn2}) from a view point of how disperse solutions are in terms of flux arrangements and biological functionality.

The similarity between two solutions $\alpha$ and $\beta$ with flux vectors $\mathbf{s}_\alpha$ and $\mathbf{s}_\beta$ respectively is conveniently quantified by their overlap. For each reaction $i$, we define
\begin{equation}
q_{\alpha\beta}(i)=\begin{cases}
1&\text{for $s_{i\alpha}<\epsilon$ and $s_{i\beta}<\epsilon$}\\
1-s_{i\alpha}&\text{if $\epsilon<s_{i\alpha}<1$ and $s_{i\beta}<\epsilon$}\\
0 &\text{if $s_{i\alpha}\geq 1$ and $s_{i\beta}<\epsilon$}\\
\frac{2 s_{i\alpha}s_{i\beta}}{s_{i\alpha}^2+s_{i\beta}^2} &\text{otherwise}
\end{cases}
\end{equation}
where $\epsilon$ is a cut-off value below which we treat fluxes as zero (results do not depend on the choice of the cut-off as long as $\epsilon$ is $10^{-5}$ or less). This parameter measures in essence the variation of flux $i$ in solutions $\alpha$ and $\beta$. For a complete flux configuration, we define
\begin{equation}
q_{\alpha\beta}=\frac{1}{N}\sum_i q_{\alpha\beta}(i)
\end{equation}
as a characterization of the ``distance'' between solutions $\alpha$ and $\beta$: the closer $q$ is to 1 (resp. 0) the more similar (resp. dissimilar) the two configurations are. Likewise, we define the overlap between the vectors $\mathbf{c}$ that correspond to the production profiles in each solution of (\ref{vn2}). The distributions of these overlaps are displayed in Figure 5.
\begin{figure}
\begin{center}
\includegraphics[width=6cm]{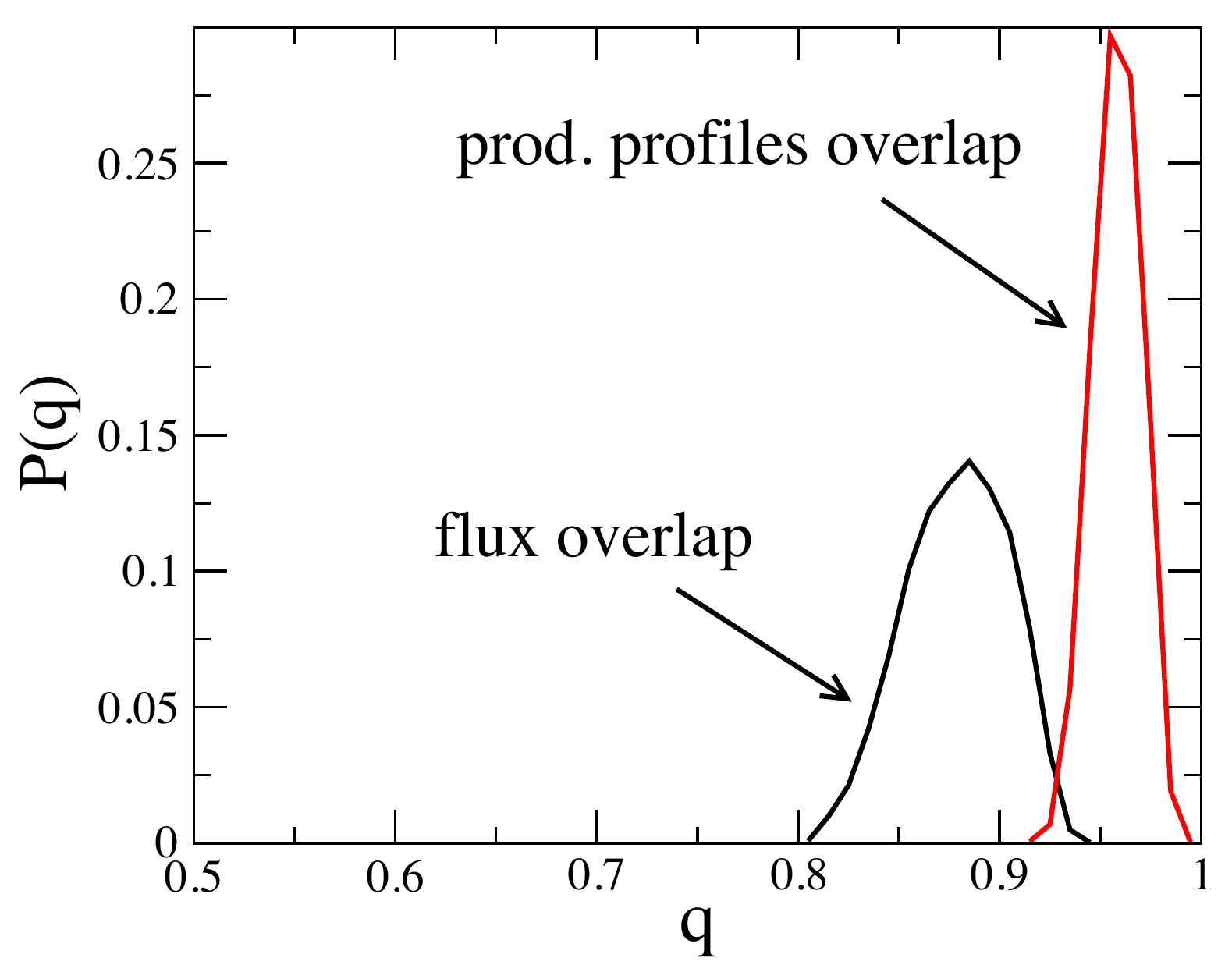}
\includegraphics[width=5.4cm]{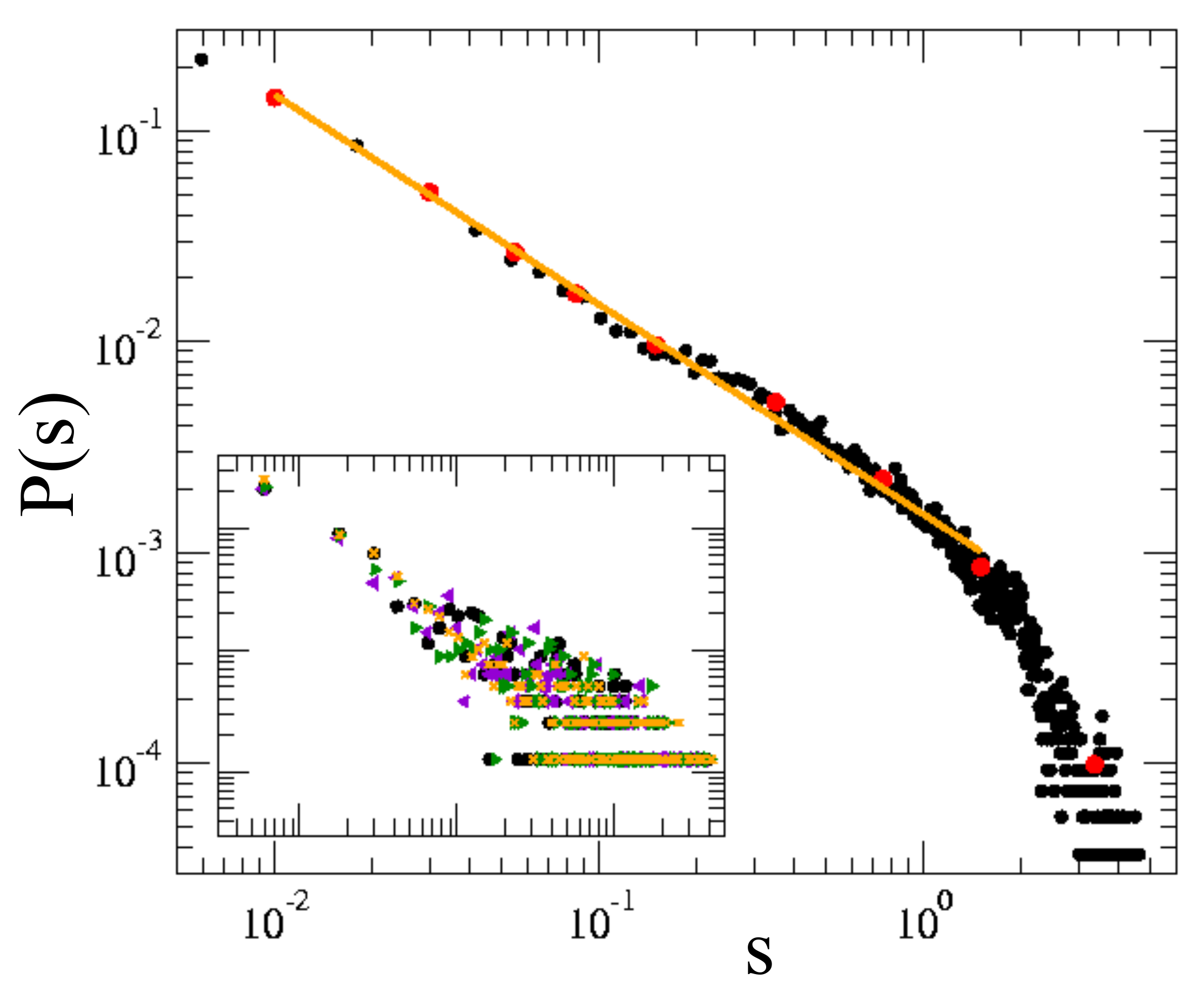}
\caption{Von Neumann-optimal states of the {\it Escherichia coli} metabolism. (Left) Distribution of the flux overlap (black) and production profile overlaps (red) over distinct solution pairs obtained from a set of 250 solutions of (\ref{vn2}). (Right) Flux distributions obtained from (\ref{vn2}). Black markers denotes the distribution of average fluxes, the orange curve corresponds to the power-law fit with exponent $-1$ and red markers correspond to a convenient binning of the data. Inset: flux distributions in various solutions (different colors). Flux units are arbitrary.}
\end{center}
\end{figure}
One sees that the distance between solutions has a broader distribution than that between production profiles, indicating that a degree of flexibility in the flux organization is allowed that nevertheless ensure a remarkable robustness in the emerging production profile. The finding of \cite{Kyoto10} that indeed the most probable output of the network is formed by the biomass constituents \cite{Feist:2010oa} is hereby fully confirmed (data now shown), as is the fact that the distribution of fluxes (see Figure 5) retains the empirically observed \cite{Almaas:2004fh} power-law shape with exponent $-1$. Single solutions however display a certain degree of variability, as can be seen again in Figure 5. 

Let us now analyze how this scenario reflects in the values of some of the recurrent physiological objective functions used in the literature (see Figure 6). We have monitored in particular: the total production of ATP (adenosine-triphosphate, the key molecular energy carrier in cells), the total consumption of NADH (nicotinamide adenine dinucleotide, the key reducing agent in electron transfer processes in metabolism, with several other important functional roles at different levels in cells), the total net flux (which is assumed to be a good proxy for enzymatic efficiency) and the glucose (GLC) uptake (measuring nutrient consumption). 
\begin{figure}
\begin{center}
\includegraphics[width=13cm]{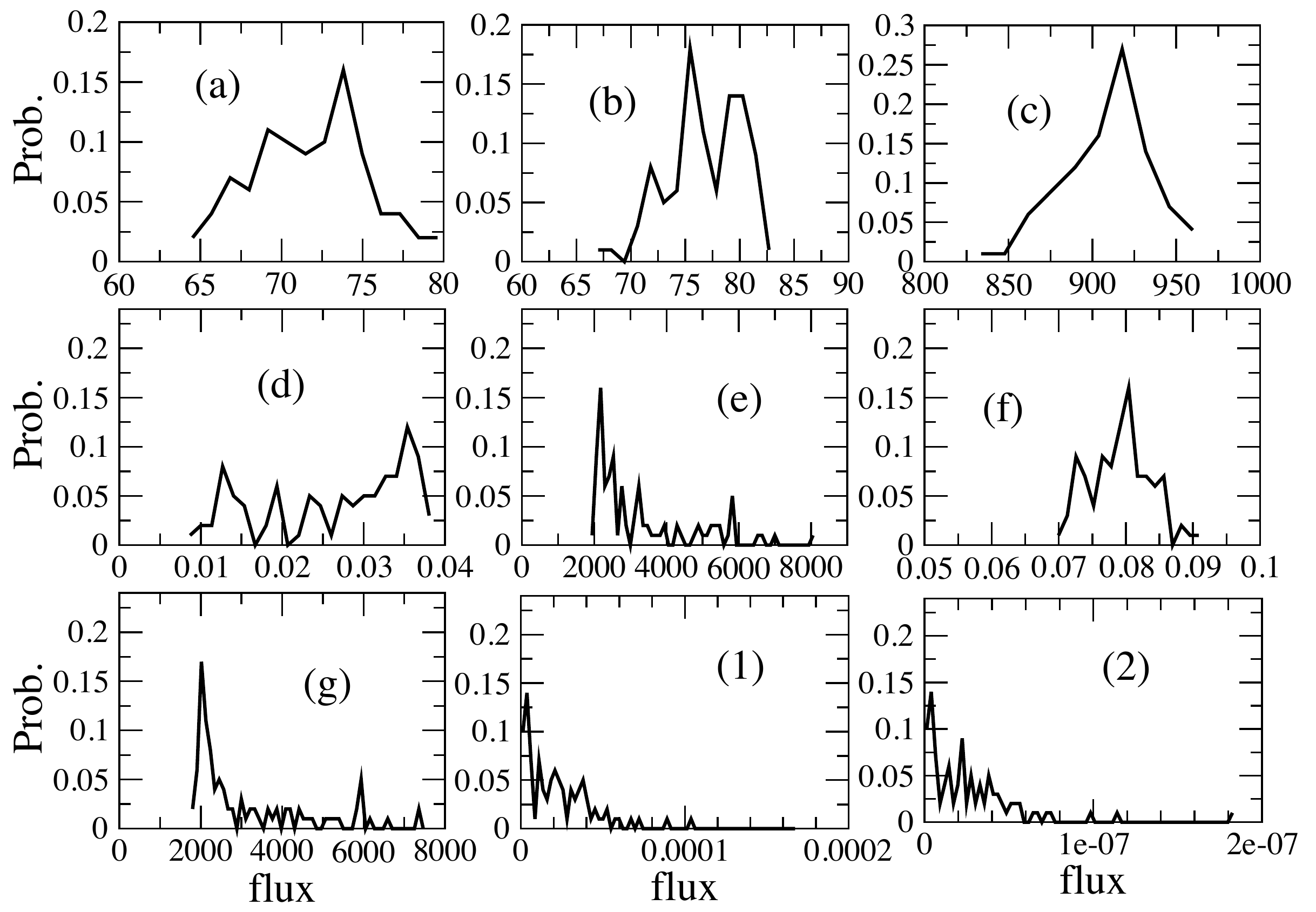}
\caption{Von Neumann-optimal states of the {\it Escherichia coli} metabolism. Probability distributions of the values of commonly employed physiological objective functions obtained from the solutions of (\ref{vn2}): (a) bare ATP yield; (b) bare NADH consumption; (c) total net flux; (d) bare glucose consumption; (e) bare NADH consumption per unit of GLC consumed; (f) bare ATP yield per flux unit; (g) bare ATP yield per unit of GLC consumed. Plots (1) and (2) display the flux pertaining to a biomass production reaction computed in a network reconstruction where such a reaction was explicitly included. This flux represents an addition over the biomass that the network spontaneously yields: (1) bare extra biomass flux; (2) bare extra biomass flux per flux unit. Flux units are arbitrary.}
\end{center}
\end{figure}
ATP production, NADH consumption and total net flux display a roughly unimodal shape, pointing to the fact that under Von Neumann's hypothesis the network is capable of tuning these functions within a relatively small range of values. Notice that the total flux is usually assumed to be minimized by cells, to account for optimal use of the available pool of enzymes. On the other hand, it is reasonable to think that cells under certain conditions maximize the ATP yield for optimal energetic efficiency. A similar picture holds for the ATP yield per flux unit, which is usually assumed to be maximized so as to guarantee optimal energetic efficiency at minimum enzyme usage. NADH consumption is instead thought to be minimized, to reduce the activity of redox processes. GLC consumptions displays instead a somewhat broader profile. This is consistent with previous results \cite{Kyoto10} and points to the fact that the emerging physiologic scenario (which appears to be robust from the energetic and enzymatic efficiency viewpoints) is indeed recoverable under widely varying GLC consumptions. In other words, glucose availability can fluctuate within a broad range without affecting optimality in the Von Neumann sense. This however does have an effect on other physiological observables like the ATP production or the consumption of NADH per unit of GLC consumed. These functions turn out to be broadly distributed as well, although with well defined peaks at low values. Finally, we have considered the effect of adding explicitly the biomass production reaction to the pool of processes encoded in the network reconstruction. Recalling that the system globally outputs biomass constituents even in absence of this additional reaction, Figure 6 shows that the extra biomass flux due to the inclusion of the additional process is essentially null. In other words, the unforced biomass output of (\ref{vn2}) by itself fully describes the cell's production capacity in the environment we selected. In summary, the physiological scenario emerging from Von Neumann's model applied to {\it Escherichia coli}'s metabolism is one in which the variability of optimal flux configurations underlies considerable robustness at the level of production profiles as well as at the level of key metabolic functions, the main exception being glucose consumption, which instead is allowed to fluctuate over a large relative range. The latter aspect contributes to robustness as well.

To conclude, we have investigated the problem of if and how forcing an extra biomass flux from the cell (i.e. forcing a non-zero value for the flux of the biomass reaction) changes the physiological scenario with respect to the reference case in which the biomass output is obtained self-consistently from the solutions of (\ref{vn2}). Figure 7 shows how the bare ATP yield and the ATP yield per flux unit (two elementary parameters to quantify the physiological state of the cell) change upon increasing the extra biomass flux.
\begin{figure}
\begin{center}
\includegraphics[width=12cm]{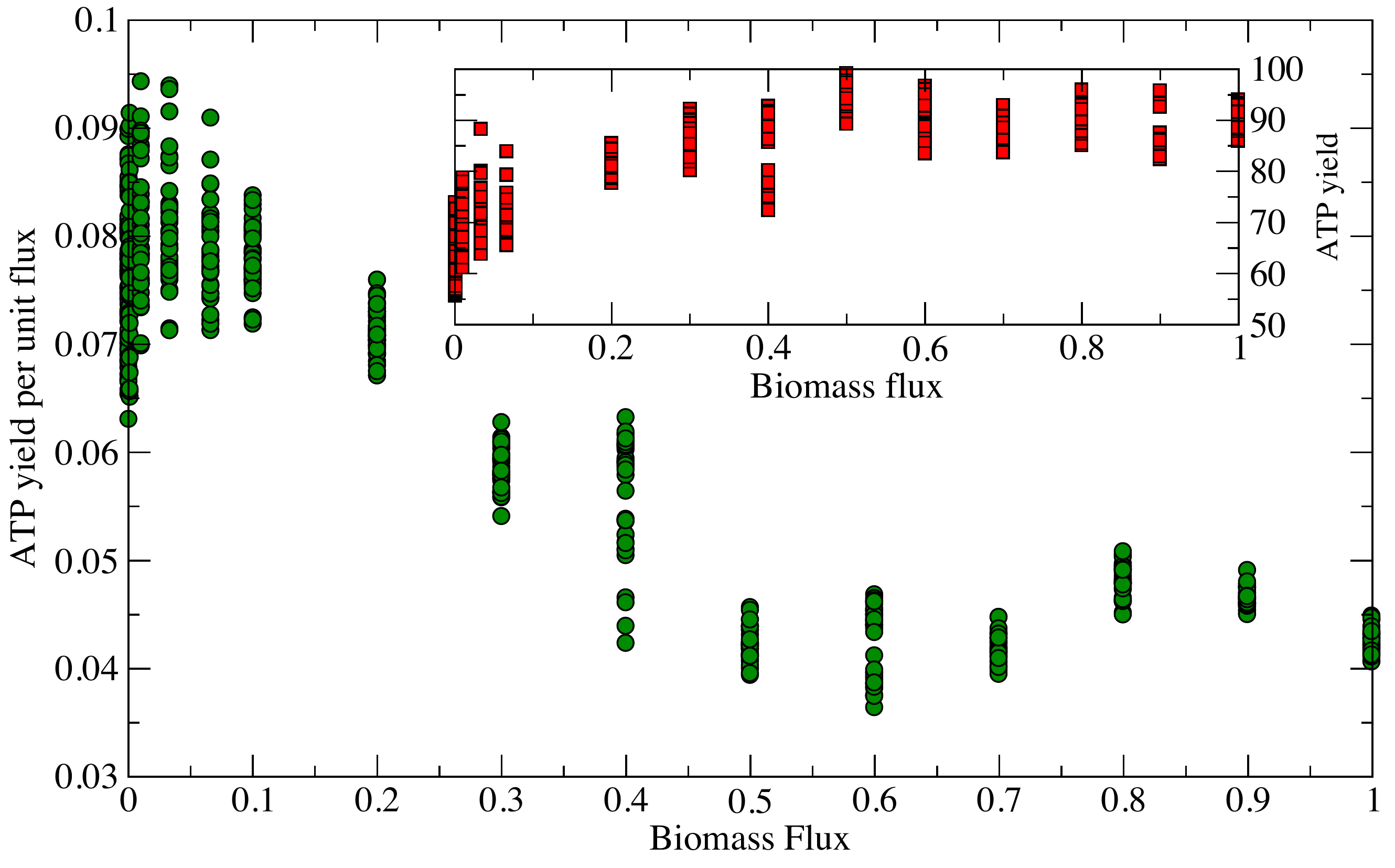}
\caption{Von Neumann-optimal states of the {\it Escherichia coli} metabolism. ATP yield (inset) and ATP yield per flux unit (main) versus extra biomass yield in states where an extra biomass flux is forced in (\ref{vn2}).}
\end{center}
\end{figure}
In first place, as should be expected, the bare ATP yield slowly increases initially  as a non-zero additional biomass flux is forced, but its value tends to saturate as the latter increases, indicating that the system has reached a maximal ATP output capacity. The relative gain in ATP yield with respect to the unforced reference state is close to 30\%. At the same time, however, the ATP yield per flux unit, which as said above quantifies energetic efficiency against enzyme usage, is actually reduced as the biomass flux increases, meaning that the cell can achieve larger biomass outputs and a gain in bare ATP yield only at the cost of a dramatic decrease in the overall metabolic efficiency (the relative loss being close to 50\%). Quite significantly, then, optimal states from the physiological viewpoint appear to be those defined by (\ref{vn2}) {\it without} any additional constraint. Similarly (data not shown) NADH consumption increases upon increasing the forced biomass flux while, by contrast, GLC consumption roughly stays constant until the extra biomass flux reaches the value $0.5$, when it starts increasing dramatically. 

These results suggest that Von Neumann's optimal states are on one hand able to provide a description of the metabolic activity of bacterial cells that is in good agreement with the empirical knowledge; on the other, they frame key physiological observables in contained ranges (implying robustness) so that forcing an improvement in the growth capacity leads to a decrease in metabolic efficiency, thereby pointing to a direct biological counterpart of Von Neumann's economic optimality. Clearly however, further work is needed in order to clarify the extent to which Von Neumann's model might be applicable in the context of metabolic networks, and above all for more complex cell types (like yeast or human cells), where regulation and kinetics may play more important roles and steady state models might turn out to be insufficient to identify the emerging physiological characteristics with a good degree of confidence.

\section{Conclusions}

Von Neumann's linear growth model addresses the very general problem of characterizing a self-sustained network of interacting input-output processes in terms of (a) the maximum achievable growth rate, (b) the pattern of activity of processes and (c) the emergent net production profile. Despite its remarkably simple rationale, this setup makes for a challenging and easily generalizable statistical mechanics problem that gives rise to a rich phenomenology which can be studied both analytically (for ensembles of random networks) and numerically (for single instances). The theory described here covers in our view only the simpler and more straightforward modifications of the original model, all guided by a chemical rather than economical intuition. From a strictly theoretical viewpoint (but possibly also with direct relevance for many applications) it would be interesting to bring them farther beyond Von Neumann's original definition, e.g. by relaxing or abandoning the assumption of constant returns to scale. Perhaps the most notable application of these ideas outside of economics (where they were originally conceived and where they form a basis for the theory of economic growth) can be located in cellular metabolism. Applying Von Neumann's growth model to cell metabolism, in our view, corresponds to appraising to what extent and in what manner do environmental, stiochiometric and thermodynamic factors limit and determine the output of a cell's biochemical machinery within a minimal physical assumption of stability. The quantitative analysis of intracellular chemical reaction networks performed along these lines has provided a physiologically sensible characterization of the metabolic capabilities of the bacterium {\it Escherichia coli} (as well as of other, simpler cell types \cite{granata}) in agreement with empirical evidence. This suggests that the function of the biochemical core of the energetics of cellular systems mirrors at least to some extent the basic notions of stability and optimality that are encoded in Von Neumann's model and possibly underlie optimal economic growth. Understanding how this picture is affected by the cross-talk between metabolism and regulation, which could be the major source of non-linearity, is perhaps the most obvious next step.

~

{\bf Acknowledgments.} We are indebted with M. Figliuzzi, D. Granata, F.A.  Massucci and V. Van Kerrebroeck and, above all, M. Marsili, C. Martelli, R. Monasson, I. Perez Castillo for long-standing, pleasant and fruitful collaborations on the issues discussed here. Our views on these matters have been crucially shaped by the insight we derived from discussion with many colleagues, including M. Avalle, W. Bialek, A.C.C. Coolen, D. De Martino, F. Di Patti, M. Di Nuzzo, S. Franz, F. Giove, T. Hwa, O. Martin, A. Pagnani, G. Parisi, F. Ricci Tersenghi, A. Seganti, D. Segr\`e, M.A. Virasoro, M. Weigt and R. Zecchina, whom we gratefully acknowledge. This work is supported by a Seed Project (DREAM) of the Italian Institute of Technology (IIT) and by the joint IIT/Sapienza Lab ``Nanomedicine''. The IIT Platform ``Computation'' is gratefully acknowledged.

\end{document}